\documentclass[aps,prb,twocolumn,showpacs,superscriptaddress,floatfix]{revtex4-1}
\usepackage{amsmath,amssymb}
\usepackage{graphicx}
\usepackage{epsfig}
\usepackage{color}
\usepackage{rotating}
\usepackage{bm}
\usepackage{setspace}
\usepackage{hyperref}

\begin{document}
\title{Molecular beam epitaxy growth and scanning tunneling microscopy study \\of TiSe$_2$ ultrathin films}
\author{Jun-Ping Peng}
\author{Jia-Qi Guan}
\author{Hui-Min Zhang}
\affiliation{State Key Laboratory for Surface Physics, Institute of Physics, Chinese Academy of Sciences, Beijing 100190, China}
\author{Can-Li Song}
\email[]{clsong07@mail.tsinghua.edu.cn}
\author{Lili Wang}
\author{Ke He}
\author{Qi-Kun Xue}
\affiliation{State Key Laboratory of Low-Dimensional Quantum Physics, Department of Physics, Tsinghua University, Beijing 100084, China}
\affiliation{Collaborative Innovation Center of Quantum Matter, Beijing 100084, China}
\author{Xu-Cun Ma}
\email[]{xucunma@mail.tsinghua.edu.cn}
\affiliation{State Key Laboratory for Surface Physics, Institute of Physics, Chinese Academy of Sciences, Beijing 100190, China}
\affiliation{State Key Laboratory of Low-Dimensional Quantum Physics, Department of Physics, Tsinghua University, Beijing 100084, China}
\affiliation{Collaborative Innovation Center of Quantum Matter, Beijing 100084, China}
\date{\today}

\begin{abstract}
Molecular beam epitaxy is used to grow TiSe$_2$ ultrathin films on graphitized SiC(0001) substrate. TiSe$_2$ films proceed via a nearly layer-by-layer growth mode and exhibit two dominant types of defects, identified as Se vacancy and interstitial, respectively. By means of scanning tunneling microscopy, we demonstrate that the well-established charge density waves can survive in single unit-cell (one triple layer) regime, and find a gradual reduction in their correlation length as the density of surface defects in TiSe$_2$ ultrathin films increases. Our findings offer important insights into the nature of charge density wave in TiSe$_2$, and also pave a material foundation for potential applications based on the collective electronic states.
\end{abstract}
\pacs{68.55.-a, 73.21.-b, 73.22.-f, 68.37.Ef}
\maketitle
\begin{spacing}{1.00}
Transition metal dichalcogenides (TMDCs) typically crystallize into layered structures via weak van der Waals attraction between adjacent layers and exhibit a variety of technologically fascinating physical properties. Like graphene, a body of distinctively promising phenomena emerges when the TMDC bulk crystals are thinned down to mono- or few-layers, which have recently attracted considerable interests in condensed matter physics and materials science.\cite{Butler2013} These phenomena include, for example, realization of the two-dimensional (2D) semiconductor with a direct band gap in the visible range,\cite{Mak2010, Zhang2014} broken parity symmetry,\cite{Cao2012, Xiao2012} pronounced spin-orbital coupling/splitting,\cite{Kosmider2013, Kormanyos2014} and extremely large exciton binding energy.\cite{Ugeda2014} The intriguing physical properties in TMDC monolayers can be employed to develop applications in optoelectronics, valleytronics, spintronics and energy storages.\cite{Butler2013, Mak2010, Zhang2014, Cao2012, Xiao2012, Kosmider2013, Kormanyos2014, Ugeda2014} Some of layered TMDCs are found to exhibit generic instabilities towards the symmetry-reducing charge density wave (CDW) and superconductivity, and therefore provide unprecedented opportunities to investigate their interplays. Parallels between TMDCs and cuprates, both of which share similar ground states, have indeed been recently claimed.\cite{Weber2011}

Titanium diselenide (TiSe$_2$), a semimetal in nature with hexagonally packed TiSe$_6$ octahedra (1\textit{T}),\cite{Bachrach1976, Cercellier2007, Rohwer2011} represents a widely studied and interesting TMDC. It undergoes a second-order phase transition to nonchiral CDW with a commensurate 2 $\times$ 2 $\times$ 2 superstructure at the CDW transition temperature $T_{\textrm{CDW}}\sim$ 200 K,\cite{DiSalvo1976} and then to chiral CDW at a slightly lower temperature.\cite{castllan2013} Yet, in spite of more than three decades of intensive experimental and theoretical endeavors, the driving force for the CDW transition remains unsettled. Upon intercalation with copper \cite{Morosan2006} or applying pressure,\cite{Kusmartseva2009} the CDW ordering melts and superconductivity develops with a critical transition temperature of several Kelvin, indicating competition between CDW and superconductivity in TiSe$_2$. Recently, self-induced topologically nontrivial and chiral superconducting phases have been predicted in pressurized TiSe$_2$ \cite{Zhu2014} and TiSe$_2$ monolayer,\cite{Ganesh2014} respectively, which might harbor the long-pursuing Majorana fermions.\cite{Ivanov2001} According to Raman spectroscopy study, $T_{\textrm{CDW}}$ can be enhanced as the thickness of mechanically exfoliated TiSe$_2$ films is reduced to nanometer scale.\cite{Goli2012} This may lead to CDW collective state-related device applications of TiSe$_2$ at room temperature. Fabrication and study of mono- or few-layer films of TiSe$_2$ are therefore especially desired.

In this work, we carry out molecular beam epitaxy (MBE) growth of TiSe$_2$ ultrathin films, and investigate the intrinsic defects, as well as CDW superstructure in the extreme 2D limit of TiSe$_2$ by using scanning tunneling microscopy (STM). Our experiments are conducted in a Unisoku ultrahigh vacuum STM system equipped with an MBE chamber for \textit{in-situ} sample preparation. The base pressure of the system is better than 1.0 $\times$ 10$^{-10}$ Torr. A nitrogen-doped SiC(0001) wafers (0.1 $\Omega\cdot$cm) is graphitized by heating to 1300$^\circ$C, which leads to a double-layer graphene terminated surface  as substrate for MBE growth of TiSe$_2$ films. High purity Ti (99.99$\%$) and Se (99.999$\%$) sources are evaporated from a homemade Ta boat and a standard Knudsen diffusion cells (CreaTec), respectively. Here the chemically inert nature of graphene ensures an atomically sharp interface between SiC and TiSe$_2$ films, which is similar to the case of MBE growth of Bi$_2$Se$_3$, FeSe and MoSe$_2$ on SiC.\cite{Zhang2014, Song2010, song2011molecular} After the MBE growth, the samples are immediately transferred into the STM head for data collection at 5.0 K. A polycrystalline PtIr tip, calibrated on the MBE-grown Ag films, is used throughout the experiments. STM topographic images are acquired in a constant current mode, with bias voltage ($V_\textrm{s}$) applied to the sample. Tunneling spectra are measured by disabling the feedback circuit, sweeping the sample voltage $V_\textrm{s}$, and then extracting the differential conductance \textit{dI/dV }using a standard lock-in technique with a small bias modulation of 10 meV at 987.5 Hz, unless other specified.

\end{spacing}
\begin{figure}[tb]
\includegraphics[width=1\columnwidth]{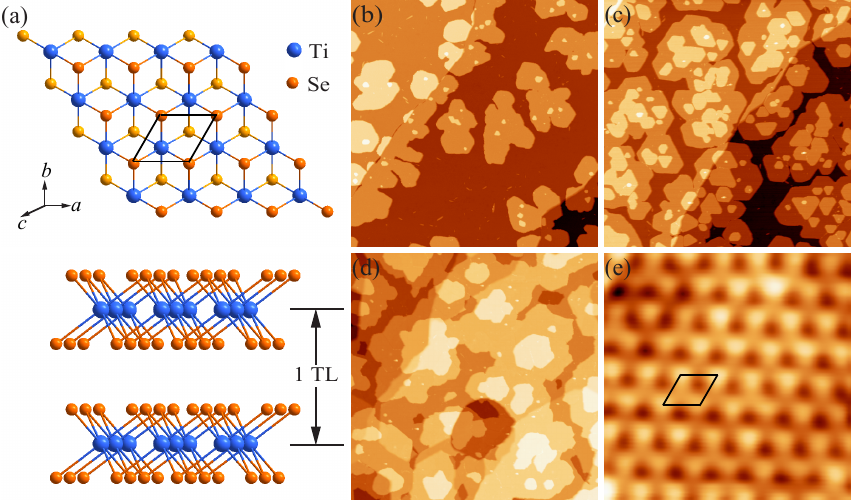}
\caption{(Color online)  (a) Schematical crystal structure of 1\textit{T}-TiSe$_2$ in the top (top panel) and side (bottom panel) views. (b)-(d) STM topographic images ($V_\textrm{s}$ = 3 V, $I$ = 30 pA, 400 nm $\times$ 400 nm) at various nominal TiSe$_2$ coverages: (a) 0.4 TL, (b) 1.0 TL, and (c) 3.8 TL. (e) Atomically-resolved STM image ($V_\textrm{s}$ = 0.2 V, $I$ = 100 pA, 2.5 nm $\times$ 2.5 nm) taken on the flat TiSe$_2$ terraces, with its unit cell marked by the black rhombus.
}
\end{figure}

As schematically illustrated in Fig.\ 1(a), a 1\textit{T}-TiSe$_2$ unit cell consists of three atomic layers  along the crystallographic (0001) direction, defining a unique triple layer (TL) of Se-Ti-Se. Within each TL, the hexagonal plane of Ti atoms is sandwiched between two Se layers, with each Ti cation coordinated octahedrally by six neighboring Se anions. These features, together with weak van der Waals interactions between adjacent TLs, bare strong similarities with those in other layered compounds, such as Bi$_2$Se$_3$, FeSe and MoSe$_2$.\cite{Zhang2014, Song2010, song2011molecular} This feature guides us to prepare TiSe$_2$ films by using the well-established MBE growth recipes for binary compounds, namely a high Se/Ti flux ratio of $\sim$ 10 and an optimal substrate temperature $T_{\textrm{sub}}$ under a condition of $T_{\textrm{Ti}}>T_{\textrm{sub}}>T_{\textrm{Se}}$ (here $T_{\textrm{Ti}}$ (1330$^\circ$C) and $T_{\textrm{Se}}$ (120$^\circ$C) are the Ti and Se source temperatures, respectively). Such conditions assure stoichiometric TiSe$_2$ films grown by MBE to be self-regulating: the Se can be incorporated only when extra Ti atoms exist on the surface of growing TiSe$_2$ films, and the growth rate of TiSe$_2$ films (0.02 TL/min) is solely determined by the $T_{\textrm{Ti}}$-controlled Ti flux.

Figures 1(b)-1(d) show the typical STM topographic images of TiSe$_2$ films grown at 200$^\circ$C, with the nominal thicknesses of 0.4 TL, 1.0 TL and 3.8 TL, respectively. Initially, TiSe$_2$ monolayer flakes, on which a small amount of tiny TiSe$_2$ islands exists as decorations, are observed and exhibit good crystallization [Fig.\ 1(b)]. With increasing coverage [Figs.\ 1(c) and 1(d)], these TiSe$_2$ flakes merge together into continuous films (especially for the bottom layer) with various thicknesses, suggesting a nearly layer-by-layer growth mode. Monolayer TiSe$_2$ nanoflakes with a corner angle of 120$^\circ$ are identified on the films [Fig.\ 1(c)], implying a hexagonal symmetry of the TiSe$_2$ films. This is confirmed by the atomically-resolved STM image taken on flat terraces [Fig.\ 1(e)]. The measured step height of 6.0 $\pm$ 0.1 {\AA}, together with the in-plane lattice constant of 3.5 $\pm$ 0.1 {\AA} extracted from Fig.\ 1(e), agrees excellently with those of (0001)-oriented TiSe$_2$ films. Note that each bright spots in Fig.\ 1(e) correspond to the topmost Se atoms.

\begin{figure}[h]
\includegraphics[width=1\columnwidth]{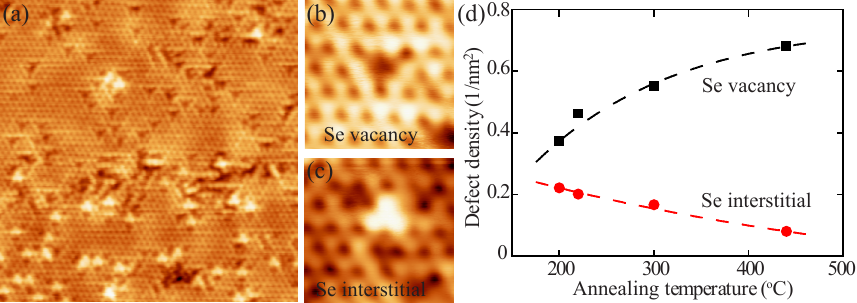}
\caption{(Color online) (a) Typical STM image of TiSe$_2$ film grown at 300$^\circ$C ($V_\textrm{s}$ = 0.24 V, $I$ = 100 pA, 17 nm $\times$ 17 nm), showing two kinds of defects: dark Se vacancies and bright Se interstitials. (b, c) Zoom-in STM images of single Se vacancy and interstitial, respectively ($V_\textrm{s}$ = 0.24 V, $I$ = 100 pA, 2 nm $\times$ 2). (d) Post-annealing temperature-dependent densities of surficial Se vacancies (black square) and interstitials (red circles) in the TiSe$_2$ films. The dashed lines are guide to eyes.
}
\end{figure}

From the atomically resolved STM images of as-grown TiSe$_2$ films [Fig.\ 2(a)], we note that there exist two dominant types of defects irrespective of $T_{\textrm{sub}}$, which appear as triangular dark depressions and bright protrusions, respectively, in the unoccupied electronic states (positive bias voltage). Such a sharp contrast, revealed in Figs.\ 2(b) and 2(c) more clearly, implies different signs of the charge states for both intrinsic defects. The dark triangular defects with their centers positioned at Se sites [Fig.\ 2(b)], are negatively charged and assigned to Se vacancies. This is consistent with previous observation in TiSe$_2$ bulk crystals.\cite{Kuznetsov2012, Hildebrand2014} As for the bright defects, they were ascribed to residual oxygen substitutions for Se sites,\cite{Hildebrand2014} which, because little oxygen is involved during the MBE growth, is not applicable to the current case.  To understand the nature of the defects, we carried out the post-annealing experiment of as-grown TiSe$_2$ films under Se flux at various temperatures, investigated the change in defect density, and plotted them in Fig.\ 2(d). As expected, the density of Se vacancies increases due to Se desorption at elevated temperature. Interestingly, the defects imaged as bright protrusions show opposite behavior: their density decreases monotonically with increasing temperature. Given this observation and the Se-rich condition used during the MBE growth, we suggest that the bright protrusions correspond probably to Se interstitials, which are expected to desorb at elevated temperature as well. If this is the case, the Se interstitials would be positively charged, which lowers the electronic energy level in their neighboring regions and consequently enhances their contrast in the STM topographies of the empty states. The assignment matches well with our observations in Figs.\ 2(a) and 2(c).

Having identified the nature of intrinsic defects in TiSe$_2$ ultrathin films, we now turn to investigate their electronic structure and CDW behavior in the 2D limit. Figure 3 represents the respective differential conductance \textit{dI/dV} spectra, which are approximately proportional to the local density of states (DOS), on defect-free single, double and five TL TiSe$_2$ films. The pronounced peaks at around 0.7 eV [Fig.\ 3(a)], observed in their bulk counterpart as well (see Fig.\ 3 in ref.\ 24), originate predominantly from Ti 3d bands.\cite{Bachrach1976, Cercellier2007, Rohwer2011, Zhu2014} A minor change ($<$ 50 meV) in peak energy positions might be caused by the dimensionality and/or electron-doping effects from the underlying n-doped graphene/SiC(0001) substrate. Remarkably, the Se 4p band-derived occupied electron DOS are considerably pushed to lower energies as the film thickness reduces, in particular in TiSe$_2$ monolayer, primarily due to the quantum confinement effects.\cite{Zhang2014} This leads to a significant depression in the population of occupied electron DOS at and near the Fermi level ($E_\textrm{F}$), as clearly revealed in Fig.\ 3. The observations indicate that the electronic structure in TiSe$_2$ ultrathin films probably have significantly changed and might differ fundamentally from that of their bulk counterpart. This comes as little surprise in terms of the emerging distinctively exotic features upon moving other TMDCs from the bulk to monolayer limit.\cite{Butler2013, Mak2010, Zhang2014, Cao2012, Xiao2012, Kosmider2013, Kormanyos2014, Ugeda2014}

\begin{figure}[h]
\includegraphics[width=1\columnwidth]{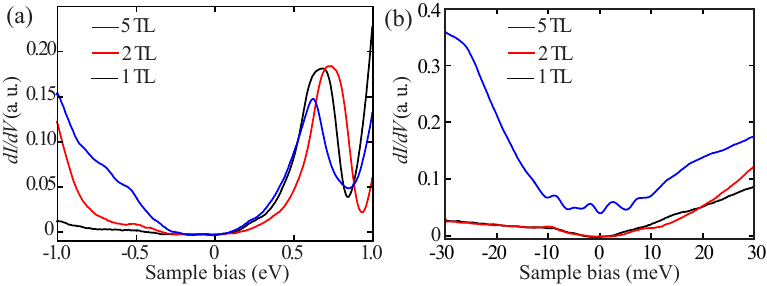}
\caption{(Color online) (a, b) Large- and small-energy-scale differential conductance \textit{dI/dV }spectra on single, double, and five layer TiSe$_2$ films. Set points are stabilized at (a) $V_\textrm{s}$ = 1.0 V and $I$ = 100 pA and (b) $V_\textrm{s}$ = 30 mV and $I$ = 100 pA, respectively. A smaller lock-in bias modulation of 0.3 meV was used for the spectra in (b).
}
\end{figure}

\begin{figure}[t]
\includegraphics[width=1\columnwidth]{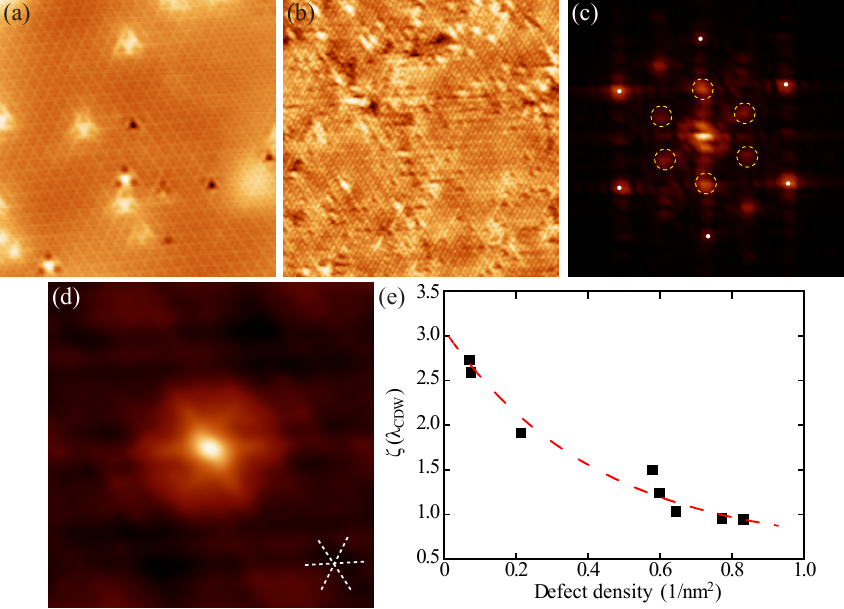}
\caption{(Color online) (a, b) STM images of CDW-related 2 $\times$ 2 superstructure in five- (a, $V_\textrm{s}$ = 2.0 V, $I$ = 100 pA, 15 nm $\times$ 15 nm) and single-layer (b, $V_\textrm{s}$ = 0.1 V, $I$ = 100 pA, 20 nm $\times$ 20 nm) TiSe$_2$ films. (c) Fourier transform pattern of STM image in (b). The white circles and yellow dashes correspond to the Bragg points and CDW 2 $\times$ 2 modulations, respectively. (d) Auto-correlation of Bragg points-filtered STM image in (b) (6 nm $\times$ 6 nm). Three white dashes mark the CDW-modulated orientations. (e) Correlation length $\zeta$ as a function of surface defect density in TiSe$_2$ films, with the red dashes as guide to eyes.
}
\end{figure}

The above-observed fundamental DOS variation may alter the Fermi surface (FS) and affect noticeably the well-known CDW phase in TiSe$_2$,\cite{DiSalvo1976} a study of which will help understand the nature of CDW in TiSe$_2$. Unexpectedly, however, our STM measurements reveal a similar commensurate 2 $\times$ 2 superstructure (originated from CDW) with a periodicity of 7.1 {\AA} $\times$ 7.1 {\AA} in TiSe$_2$ ultrathin films down to single TL, as depicted in Figs.\ 4(a) and 4(b). This is more clearly visible in the 2D Fourier transform image [Fig.\ 4(c)]. The result resembles greatly with those observed in TiSe$_2$ bulk crystals,\cite{Ishioka2010p, Ishioka2010} although the CDW strength might exhibit a substantial difference. As a result, our observations indicate that the traditional FS nesting scenario may be not applied to the CDW in TiSe$_2$,\cite{Gruner1988} as the aforementioned change in electron DOS most likely breaks the strict FS nesting condition that gives rise to CDW. Alternatively, the band-type Jahn-Teller mechanism has been proposed to be responsible for CDW in TiSe$_2$.\cite{Rossnagel2002} The starting point of this model is a band structure with substantial Se 4p-Ti 3d overlapping, which becomes worse as the Se 4p valance band shifts downwards with reducing film thickness. More significantly, in this model, a gap opening is commonly accompanied but rarely observed in the low-temperature CDW phase of TiSe$_2$ bulk and ultrathin films. Thus, it seems inappropriate to assign Jahn-Teller origin to the CDW transition in TiSe$_2$, although it needs more experimental and theoretical endeavors to wholly rule out this mechanism. Finally, we consider the possible electron-hole excitonic insulator scenario. In this scenario, three prerequisites should be generally satisfied: (i) low excess carrier concentration, (ii) large exciton binding energy, and (iii) long scattering lifetime.\cite{Rossnagel2010} The conditions (i) and (iii) are naturally fulfilled given the suppressed electron DOS near $E_\textrm{F}$ [Fig.\ 3] and high-quality films investigated here. The low electron DOS and reduced dimensionality effects can increase the exciton binding energy in TiSe$_2$ ultrathin films owing to the reduced screening, thus condition (ii) is also satisfied. Therefore our experimental observations are compatible with the electron-hole excitonic insulator mechanism for CDW formation in TiSe$_2$. The enhanced excitonic binding energy with reducing film thickness can enhance $T_{\textrm{CDW}}$, in line with the recent Reman measurements of exfoliated TiSe$_2$ films.\cite{Goli2012} A further theory and experiments (e.g. by ARPES technique) of the TiSe$_2$ films in the extreme 2D limit would eventually pin down the driving force of CDW.\cite{Taraphder2011, Koley2014}

To provide a deeper insight into CDW mechanism in TiSe$_2$, we analyze the commensurate 2 $\times$ 2 superstructure dependence on the density of surface defects, which may effectively reduce the scattering lifetime. Evidently, the CDW superstructure appears a little blurred and gets patched in monolayer TiSe$_2$ film where a large number of defects exist [Fig.\ 4(b)]. It is in contrast with that in clean 5 TL films [Fig.\ 4(a)]. In order to understand this behavior quantitatively, we have tried to measure the correlation length $\zeta$ of these CDW patches (or the effective CDW range) as a function of defect density.\cite{Arguello2014, Brun2010} To do so, we filter the atomic Bragg peaks out from the STM image, and calculate their autocorrelation image. The result is shown in Fig.\ 4(d). We then measure the intensities of three line cuts along all the three CDW-modulated orientations (three marked dashes) to extract the correlation length $\zeta$ as the full width at half maximum (FWHM) in the averaged line cut. Figure 4(e) plots the defect density-dependent correlation length $\zeta$. It is evident that the correlation length $\zeta$ decreases monotonically with increasing defect density, indicating the detrimental role of the intrinsic defects in the long-range coherence of CDW order in TiSe$_2$. The results not only support the electron-hole excitonic insulator mechanism for CDW, but also suggest that ultrathin TiSe$_2$ films with much lower defect density is essential for future realization of CDW-based application in TiSe$_2$.

The successful MBE growth of TiSe$_2$ ultrathin films down to monolayer thickness demonstrates an alternative approach to fabricate mono- and few-layer TMDC materials. Our STM measurements reveal three pieces of important information about TiSe$_2$ ultrathin films. First, we have identified two dominant kinds of Se vacancy and interstitial defects, with their concentrations critically dependent on substrate temperature $T_{\textrm{sub}}$. Second, by \textit{dI/dV} spectra we have demonstrated that the band structures of TiSe$_2$ ultrathin films may differ fundamentally from those of their bulk counterpart. Third, the observation that CDW persists down to TiSe$_2$ monolayer favors the excitonic insulator mechanism for CDW in TiSe$_2$.

\begin{acknowledgments}
This work was financially supported by National Science Foundation and Ministry of Science and Technology of China. All STM images were processed by Nanotec WSxM software.\cite{horcas2007wsxm}
\end{acknowledgments}
%
\end{document}